\documentclass[12pt,preprint]{aastex}
\usepackage{graphicx}

\slugcomment{To appear in the Astrophysical Journal}
\shorttitle{Neon Insights from Old Solar X-rays}
\shortauthors{Drake}
\begin{document}

\title{NEON INSIGHTS
FROM OLD SOLAR X-RAYS: A PLASMA TEMPERATURE
DEPENDENCE OF THE CORONAL NEON CONTENT}

\author{Jeremy J.~Drake\altaffilmark{1}}
\affil{$^1$Smithsonian Astrophysical Observatory,
MS-3, \\ 60 Garden Street, \\ Cambridge, MA 02138, USA}
\email{jdrake@cfa.harvard.edu}

\begin{abstract}
An analysis using modern atomic data of fluxes culled from the literature for 
O~VIII and Ne IX lines observed in solar active regions
by the P78 and Solar Maximum Mission satellites 
confirms that the coronal Ne/O abundance ratio varies by a factor 
of two or more, and finds an increase in Ne/O with increasing active region plasma temperature.   The latter is reminiscent of evidence for 
increasing Ne/O with stellar activity in low-activity coronae that reaches a ``neon saturation'' in moderately active stars at approximately twice the historically 
accepted solar value of about 0.15 by number.  We argue that neon saturation represents the underlying stellar photospheric compositions, and that low activity coronae, including that 
of the Sun, are generally depleted in neon.  The implication would be that the 
solar Ne/O abundance ratio should be revised upward by a factor of about two 
to $n($Ne$)/n($O$)\sim 0.3$.  Diverse observations of neon in the local cosmos provide some support for such a revision.  Neon 
would still be of some relevance for
reconciling helioseismology with solar models computed using 
recently advocated chemical mixtures with lower metal content.

%
\end{abstract}


\section{Introduction}
\label{s:intro}

Interest in Neon, the newcomer, has been strongly piqued recently by
two possibly related puzzles concerning the value of the Ne abundance
in the Sun and stars.  In a study of 21 mostly magnetically active
stars observed by the {\it Chandra} high resolution X-ray
spectrometers, \citet{Drake.Testa:05} found Ne/O abundance ratios to
be consistently higher by a factor of 2.7 than the 
recommended solar value of the time, Ne/O=0.15 by number\footnote{Throughout we 
use the notation Ne/O to refer to the ratio by number of Ne to O nuclei, rather than
the implicit logarithmic abundance ratio sometimes used.}.  The latter 
value is traceable to the assessment of \citet{Anders.Grevesse:89}.

\citet{Drake.Testa:05} suggested that the same ratio might also be
appropriate for the Sun and help solve the ``Solar Model Problem''.
Models employing a recently advanced solar chemical
composition based on 3-D non-LTE hydrodynamic photospheric modelling
\citep{Asplund.etal:05} led to predictions of the depth of the
convection zone, helium abundance, density and sound speed in serious
disagreement with helioseismology measurements
\citep{Basu.Antia:04,Bahcall.etal:05}.  The \citet{Asplund.etal:05}
mixture contained less of the elements C, N, O and Ne that are
important for the opacity of the solar interior by 25-35~\%\ compared
to earlier assessments
\citep[e.g.][]{Anders.Grevesse:89,Grevesse.Sauval:98}.
\citet{Antia.Basu:05} and \citet{Bahcall.etal:05c} suggested the uncertain solar
Ne abundance might be raised to compensate.  The estimated 
Ne/O ratio required was in agreement with that found in stellar coronae by
\citet{Drake.Testa:05}.

More detailed investigations of helioseismology models now suggest that 
raising the Ne abundance alone cannot fully reconcile models and 
observations \citep{Delahaye.Pinsonneault:06,Lin.etal:07,Delahaye.etal:10}. Nevertheless, the Ne abundance still remains an important individual ingredient, not only for the Sun but also for nucleosynthesis and galactic chemical evolution. 

Ne exhibits no lines in the visible light spectra of late-type stars and is not retained in meteorites.  Consequently, 
the solar abundance is based largely on transition region and
coronal lines, and energetic particle measurements, supplemented with
local cosmic estimates \citep[e.g.\ ][]{Meyer:85,Anders.Grevesse:89,Asplund.etal:09}.  Two studies of transition region EUV and coronal X-ray lines find
consistency with the ratio Ne/O=0.15
\citep{Young:05,Schmelz.etal:05b} and conclude that this ratio
represents that of the bulk of the Sun.  However, based on UV lines observed during a flare \citet{Landi.etal:07} 
obtained an absolute abundance Ne/H a factor of 1.9 larger than the
\citet{Asplund.etal:05} value.  
Subsequently, the latter authors have revised their recommended Ne and O abundances back up toward older values by 0.13 and 0.07~dex, respectively \citep{Asplund.etal:09}, raising their favoured absolute logarithmic Ne abundance to Ne/H$=7.97\pm 0.1$ and 
 Ne/O ratio to 
0.17---just within formal agreement with the \citet{Landi.etal:07} Ne abundance, but still somewhat short of the \citet{Drake.Testa:05} Ne/O ratio.

Another important clue to the solar neon content is provided by past coronal X-ray
Ne/O measurements.  Results reviewed by  \citet{Drake.Testa:05} showed considerable
scatter, and an extensive study of active regions has demonstrated
variations in the coronal Ne/O ratio by factors of more than 2
\citep{McKenzie.Feldman:92}.  Such variations indicate coronal
fractionation of Ne relative to O by mechanisms that are not yet firmly 
identified or understood, and cast serious doubt on the validity of
conclusions based on a simple average of observed values to infer the 
Ne content of the Sun.  

The variation in observed solar Ne/O ratios motivates re-examination of
existing observations using modern 
atomic data for further insights into the behaviour of Ne in the
solar outer atmosphere.  Unfortunately, some earlier studies of Ne in
the solar X-ray spectrum did not list spectral line fluxes, rendering
re-analysis somewhat difficult.  Observations made between 1979 March 23 and November 29 by the SOLEX
spectrometer on the US Department of Defense P78-1 satellite analysed
by \citet{McKenzie.Feldman:92}, and the recent study by
\citet{Schmelz.etal:05b} of {\it Solar
Maximum Mission} (SMM) Flat Crystal Spectrometer (FCS) observations obtained between 1986 May 20 and 1987 December 18 are two 
exceptions that form the basis of this
work.

\section{Analysis}
\label{s:anal}

\subsection{Methods and Atomic Data}
\label{s:methods}

The analysis presented here is similar to that originally applied to
Ne and O lines by \citet{Acton.etal:75}, and duplicated in successive
papers including \citet{McKenzie.Feldman:92} and
\citet{Schmelz.etal:05b} (see also \citealt{Drake.Testa:05}): Ne/O
abundance ratios are inferred from observed ratios of the intensities
of the transitions Ne~IX $\lambda 13.47$ $1s2p\, ^1P_1 \rightarrow
1s^2\, ^1S_0$ and O~VIII $\lambda 18.97$ $2p\, ^2P_{3/2,1/2}
\rightarrow 1s\, ^2S_{1/2}$ and comparison with theoretical
predictions.

Line intensities were taken from Table~2 of
\citet{McKenzie.Feldman:92} and Table~2 of \citet{Schmelz.etal:05b}. 
One possible complication to the analysis of the 
Ne~IX $\lambda 13.47$ line is the potential presence of Fe~XIX 
blends \citep[see, e.g.][]{Ness.etal:03}; both \citet{McKenzie.Feldman:92} and \citet{Schmelz.etal:05b} take pains to note this and assess that for their non-flaring 
spectra such blends are negligible.  This assessment is verified below
when specific account of the Fe~XIX contribution to Ne~IX $\lambda 13.47$ 
is taken.

The theoretical Ne~IX/O~VIII line ratio has significant
temperature dependence and derivation of an abundance ratio from a measured flux ratio requires an estimate of temperature \citep[see] [for a more temperature-insensitive ratio using both Ne~IX and Ne~X lines]{Drake.Testa:05}.
\citet{McKenzie.Feldman:92} utilised the ratio of
Fe~XVIII
$\lambda 14.21$ $2p^5\, ^2P_{3/2} - 2p^4(^1D)3d\, ^2D_{5/2},
^2P_{3/2}$ and Fe~XVII $\lambda 15.01$ $2p^6\, ^1S_0 - 2p^53d\, ^1P_1$ as a temperature diagnostic.  They list the line fluxes
which we use here to re-derive temperatures using the atomic data noted below.
\citet{Schmelz.etal:05b} used the same Fe~XVIII lines
combined with Fe~XVII $\lambda 16.68$ $2p^6\, ^1S_0
 - 2p^53s\, ^3P_1$ but omit measured fluxes and only list their
derived temperatures which we employ directly.   Questions have
been raised in the past regarding possible optical depth of 
Fe~XVII $\lambda 15.01$ due to resonance scattering: 
\citet{Brickhouse.Schmelz:06} have recently concluded the line is consistent with being optically thin in FCS observations of active regions, and attribute 
apparent quenching to Fe~XVI blends in the neighbouring Fe~XVII~$\lambda 15.26$ 
line often used as an optically thin comparison.

Spectral line fluxes were analysed using the
PINTofALE\footnote{PINTofALE is freely
available from http://hea-www.harvard.edu/PINTofALE/} IDL\footnote{Interactive
Data Language, Research Systems Inc.} software
\citep{Kashyap.Drake:00}, employing collisional excitation rates and
energy levels from the CHIANTI database v6.0.1 \citep[][and references
therein]{Landi.etal:06,Dere.etal:97}, together with the ionization fractions as 
function of temperature from \citet{Bryans.etal:09}.  The
sensitivity of the results to the adopted atomic data was also investigated.

\subsection{Ne/O Line Ratios}
\label{s:linerats}

\subsubsection{Isothermal Analysis}
\label{s:iso}

We first illustrate in Figure~\ref{f:linerats} the
\citet{McKenzie.Feldman:92} and \citet{Schmelz.etal:05b} Ne/O line
ratios (in photon units) as a function of the isothermal plasma
temperature.  
Uncertainties in the temperatures, both from \citet{Schmelz.etal:05b} and derived here using the \citet{McKenzie.Feldman:92} Fe~XVII and Fe~XVIII line fluxes and 
measurement uncertainties, amount typically to no more than 0.05-0.1~dex and are omitted for clarity.  
The \citet{McKenzie.Feldman:92} and \citet{Schmelz.etal:05b} Ne/O flux ratios
are consistent with one another and follow a steep, approximately
linear, trend of increasing ratio with increasing temperature.   

We now examine whether the data simply reflect the temperature dependence of the Ne~IX/O~VIII line ratio and are consistent with a single Ne/O abundance ratio.  To do this, we first compare the observed flux ratios with the theoretical ratio computed as a function of the isothermal plasma temperature.  This comparison assumes the individual observations also correspond to isothermal plasma; this is strictly not likely to be the case and we return to this later in Section~\ref{s:dem}.  

The theoretical line intensity ratio depends linearly on the assumed Ne/O abundance ratio, and a single theoretical intensity ratio vs temperature locus was found to be a poor match to the observed data for any single adopted Ne/O abundance ratio.   
In Figure~\ref{f:linerats} we illustrate this by showing two theoretical isothermal intensity ratio curves corresponding to two different Ne/O abundance ratios.  These curves are analogous to both the upper panel in Figure~1 of \citet{Acton.etal:75}, and the curves bounding
the line flux ratios in Figure~10 of \citet{McKenzie.Feldman:92}. 
The Ne/O abundance ratio for the lower curve (Ne/O=0.12) is such that the {\em minimum} theoretical photon intensity ratio reached matches the {\em minimum} observed ratio, excluding
the most extreme observed point.  The abundance ratio for the upper curve (Ne/O=0.17) is such that the {\em maximum} theoretical photon intensity ratio reached matches the {\em maximum} observed ratio, again excluding
the most extreme observed point.  Note that there are still observed photon intensity ratios that lie below the lower theoretical ratio curve and one that lies slightly above the upper one: in principle, these could be made to fit within the two curves were there to be some source of additional error in the derived plasma temperatures which allowed the points to be shifted arbitrarily to the left or right.  In this way, the two curves represent the {\em minimum possible spread} in Ne/O abundance ratio (Ne/O=0.12 and 0.17) that can match the data.  The separation of the two curves is much greater than the statistical errors in the data, and we conclude that the observed intensity ratios cannot be explained by a single Ne/O abundance ratio.  

The trend of the observed intensities with temperature is also much steeper than the theoretical ratio. If the observed intensities are from isothermal plasma, this  indicates
a trend of increasing Ne/O abundance ratio with temperature.


%

\subsubsection{DEM Analysis}
\label{s:dem}

The fields of view of the SOLEX and FCS instruments from which the observations analysed here were obtained were 1 arcmin and 15 arcsec across, respectively \citep{McKenzie.Feldman:92,Schmelz.etal:05b}.  It is likely than in regions of this size the observed plasma is not isothermal.   Any deviation from isothermality tends to flatten out the theoretical line intensity ratio curves.  We investigate this quantitatively using a model continuous differential emission measure distribution (DEM). 


For the DEM model, we adopted the form
$\Phi(T)=n_e^2(T)\frac{dV(T)}{dT}$, where $V$ is the
volume occupied by material with electron density $n_e$ at temperature
$T$.   $\Phi(T)$ was approximated by a combination of two power laws bridged 
by a constant flat top:
\begin{equation}
\Phi(T)=\phi_\alpha T^\alpha+\phi_\beta T^\beta + {\rm rect}((T_{max}-T)/\delta T), 
\end{equation}
where the DEM peaks at a temperature between $T=T_{max}\pm \delta T/2$,
\begin{displaymath}
\phi_\alpha=\begin{array}{ll}
1/T_{max}^\alpha & \mbox{for $T\leq T_{max}-\delta T/2$} \\
0 & \mbox{for $T > T_{max}-\delta T/2$}
\end{array}
\end{displaymath}
\begin{equation}
\phi_\beta=\begin{array}{ll}
0 & \mbox{for $T \leq T_{max}+\delta T/2$} \\
1/T_{max}^\beta & \mbox{for $T > T_{max}+\delta T/2$} ,
\end{array}
\end{equation}
and ${\rm rect}(T)$ is the rectangular function
\begin{displaymath}
{\rm rect}((T_{max}-T)/\delta T)=\begin{array}{ll}
 1 & \mbox{for $|T -T_{max}| < \delta T/2$} \\
0 & \mbox{for $|T - T_{max}| \geq \delta T/2$}.
\end{array}
\end{displaymath}
Here, $\alpha$ and $\beta$ are constants describing the steepness of
the rise of the DEM for $T < T_{max}$ and its subsequent 
decay for $T > T_{max}$.  Constant conductive loss models
and quasi-static, constant cross-section uniformly heated loop
models have the well-known power law slope $\alpha =3/2$ in the case
when coronal structures are small compared to the pressure scale
height \citep[e.g.][]{Jordan.etal:87}.  Observations indicate that
active regions can be characterized by much steeper $\Phi(T)$;
e.g.\ \citet{Drake.etal:00} found $\alpha\sim 4$ for the brightest solar
active regions and the coronae of the intermediate activity stars
$\xi$~Boo~A and $\epsilon$~Eri.  

The observed line intensity for a transition $ij$ in species $X$ with an abundance $A$ in a plasma with an DEM $\Phi_{T_{max}}(T)$ characterized by a peak temperature $T_{max}$ can be written 
\begin{equation}
F(T_{max})_{ij}=A\int_{0}^\infty \varepsilon_{ij}(T) \Phi(T)\; dT,
\label{eq:flux}
\end{equation}
where $\varepsilon_{ij}$ is the effective line emissivity.   We can use theoretical line fluxes as a function of the characteristic temperature, $T_{max}$, calculated 
using Equation~\ref{eq:flux}  in a similar way to the use of isothermal fluxes described above in Section~\ref{s:iso}.  As an illustrative case, we adopted 
 $\alpha=3, \beta=5$, with a small flat-topped 
maximum of width $\delta T=0.1$ in $\log T$.  Using this model DEM and the 
 \citet{McKenzie.Feldman:92} Fe~XVII and FeXVIII line intensities, we computed the 
DEM peak temperature, $T_{max}$, corresponding to all the different observations, as for the isothermal case.  The \citet{McKenzie.Feldman:92} Ne/O line intensity ratios are illustrated as a function of this $T_{max}$ as grey points in Figure~\ref{f:linerats}.  Overlaid are two dashed curves representing the 
multi-thermal DEM theoretical Ne~IX/O~VIII line intensity ratios as a function of the DEM peak temperature $T_{max}$.   
These two curves 
correspond to two Ne/O abundance ratios (Ne/O=0.09 and 0.22) chosen in the same way as for the isothermal case described in Section~\ref{s:iso}: these dashed curves are the multi-thermal equivalent of the isothermal ones illustrated as solid curves.  They represent the minimum spread in Ne/O that can explain the observations and demonstrate that under multi-thermal conditions the data cannot be explained by a single Ne/O abundance ratio.

The multi-thermal intensity ratio curves are relatively insensitive to the adopted values of $\alpha$, $\beta$ and $\delta T$: 
as slopes $\alpha$ and $\beta$ tend toward much higher
values, the curve approaches the isothermal case; for much shallower slopes
and larger values of $\delta T$, 
the curve simply becomes flatter, tending toward the constant ratio of 
line emissivities integrated over all temperatures.   
The smoother multi-thermal theoretical intensity ratios require a much larger spread in
Ne/O abundance ratio than the isothermal ones.  
Again, the observed Ne/O intensities rise much more steeply with 
peak DEM temperature than the theoretical curve, indicating a general
trend of increasing Ne/O abundance ratio with plasma temperature.
 

The conclusion that the active region Ne/O line ratios
are inconsistent with a single Ne/O abundance ratio was made earlier
by \citet{McKenzie.Feldman:92}, who found Ne/O 
varied by a factor of 2.4.  What is new here is the trend of
increasing Ne/O with increasing plasma temperature.

We note in passing that indications of Ne/O abundance variations of a factor of about 2, 
consistent with those
found here and by \citet{McKenzie.Feldman:92}, are also apparent in the 
earlier analysis of Ne, O and Fe
lines in some SMM FCS spectra by \citet{Strong.etal:88}.   
Their Figure~4 shows observed Ne/O intensity ratios as a function of plasma temperature that lie between theoretical curves spanning Ne/O abundance ratios of 0.33 to 0.17.  That work was focused on temperature diagnostics, and the authors drew attention to probable abundance errors but not variations.



\subsection{Ne/O Abundance Ratios}
\label{s:abunrats}

\begin{figure}
\begin{center}
\includegraphics[width=0.80\textwidth]{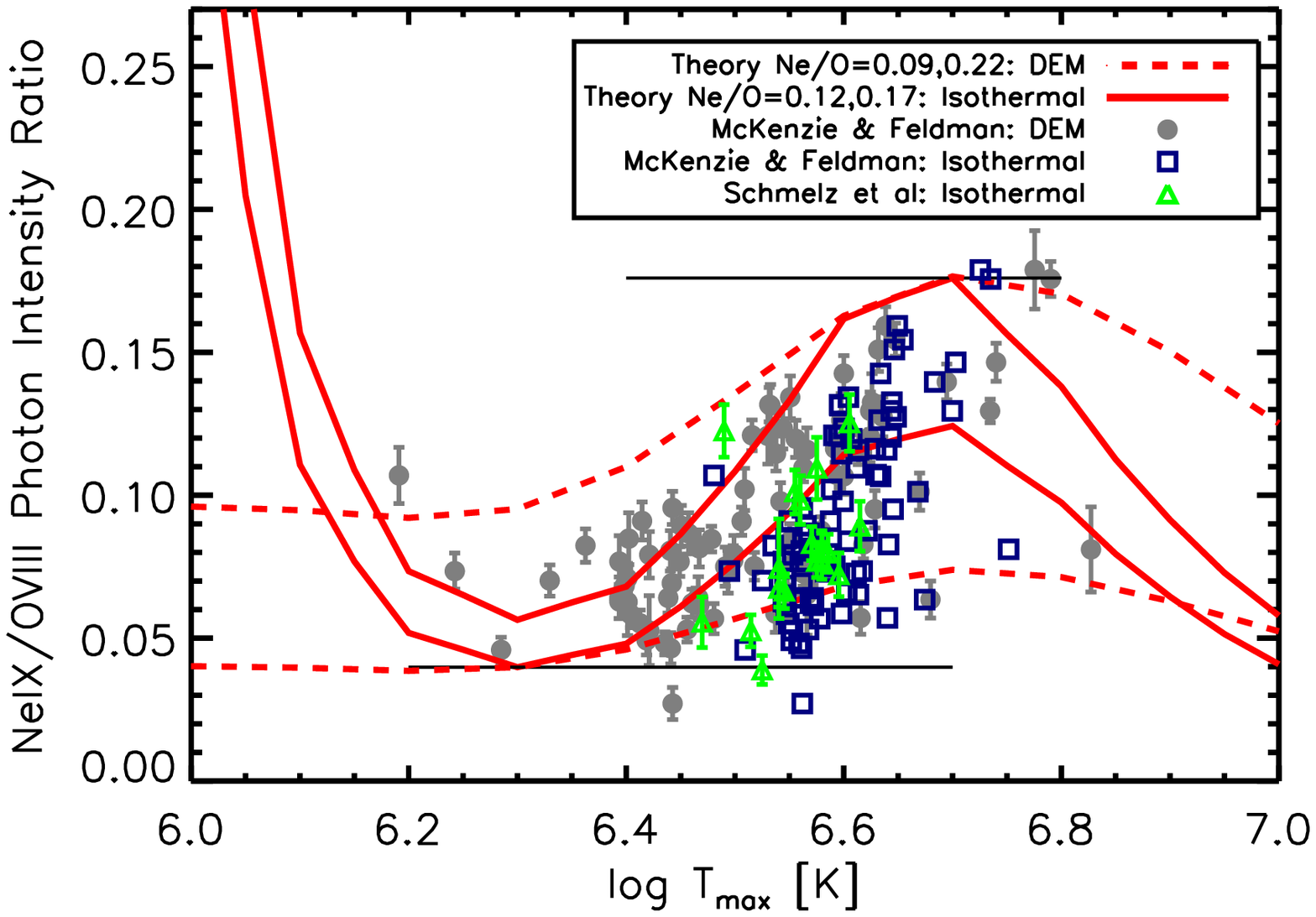}
\end{center}
\caption{Theoretical ratios of photon fluxes from the transitions 
O~VIII $2p\, ^2P_{3/2,1/2}
\rightarrow 1s\, ^2S_{1/2}$ (18.97~\AA ) and  Ne~IX $1s2p\, ^1P_1
\rightarrow 1s^2\, ^1S_0$ (13.45 \AA ) as a function of both
isothermal plasma temperature, $T$, and model DEM peak temperature
$T_{max}$, compared with observations of solar active regions 
analysed by \citet{McKenzie.Feldman:92}
and \citet{Schmelz.etal:05b}.    Each pair of solid (isothermal) and dashed 
(DEM) curves corresponds to the minimum spread in Ne/O abundance ratios that 
can possibly bracket the observations---the lower limit to the variation
in observed Ne/O where the lowest flux ratios originate from regions with 
temperature of minimum theoretical ratio, and vice versa.   These limits are denoted by black solid horizontal lines.  If regions are isothermal,
the minimum range is 42\%, from Ne/O=0.12 (lower solid curve) to 0.17 (upper solid curve) by number; for a typical DEM temperature structure (see text), the range is more than a factor of two, from Ne/O=0.09 (lower dashed curve) to 0.22 (upper dashed curve).  For clarity, error bars for \citet{McKenzie.Feldman:92} intensity ratios are only plotted on the grey points with DEM-based temperatures.
}
\label{f:linerats}
\end{figure}

\begin{figure}
\begin{center}
\includegraphics[width=0.80\textwidth]{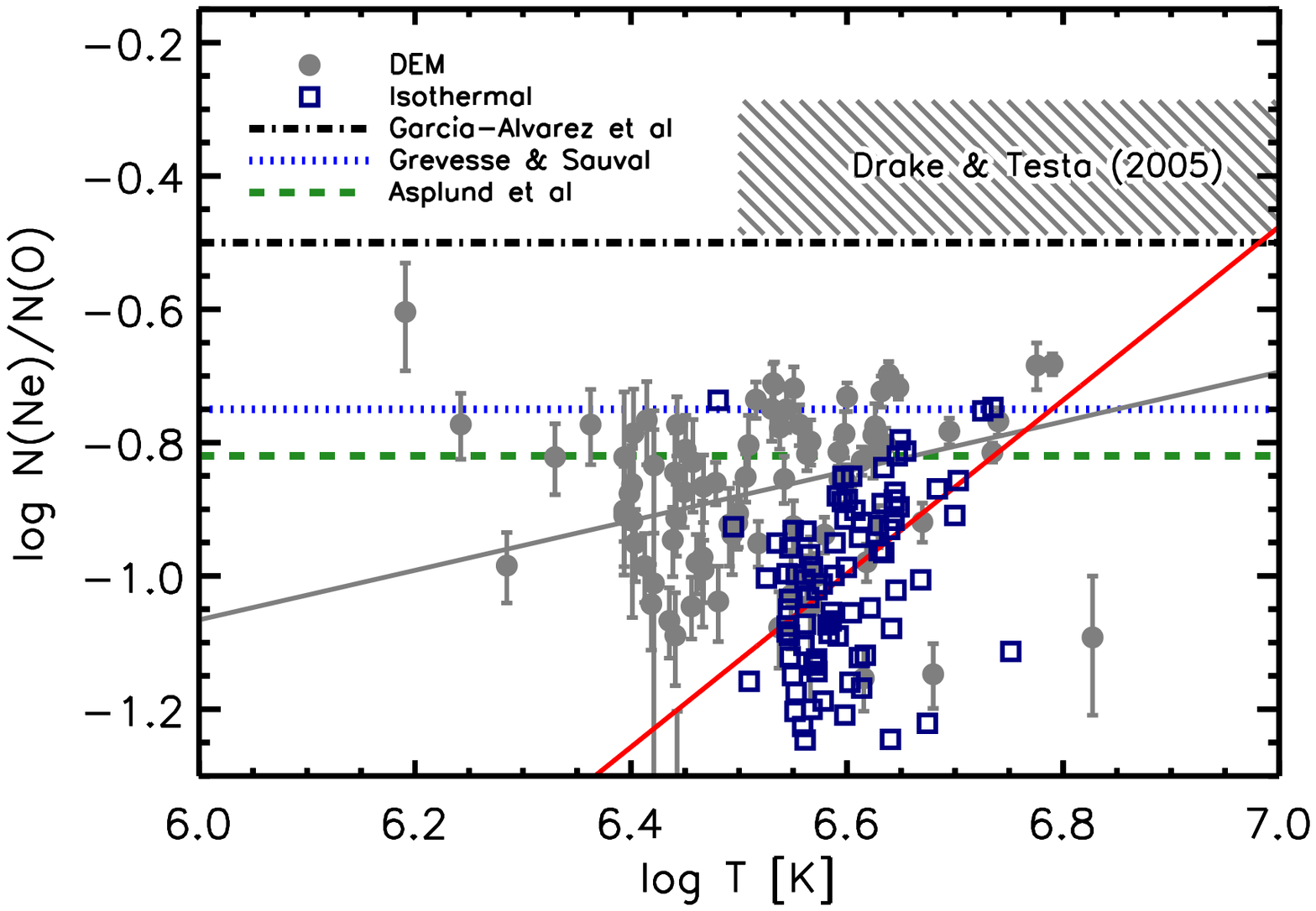}
\end{center}
\caption{Ne/O abundance ratios derived from the
  \citet{McKenzie.Feldman:92} solar active region line fluxes for both
  isothermal temperatures and using the model DEM (see text), compared
  with the recommended solar Ne/O ratio of \citet{Asplund.etal:05},
  the ``superseded'' solar ratio of \citet[][the value recommended by \citealt{Asplund.etal:09} lies in the middle of these two]{Grevesse.Sauval:98}, the 
  $\pm1 \sigma$ range 
  of values found for active stellar coronae by
  \citet{Drake.Testa:05}, and the mean for a sample of stellar coronae culled from  
  the literature by \citet{Garcia-Alvarez.etal:09}.  Linear fits
  to the derived logarithmic Ne/O abundance ratios are illustrated by straight solid lines.}
\label{f:abuns}
\end{figure}

The Ne/O abundance ratio for each of the observed 
line intensity ratios was obtained
for isothermal and DEM cases using the theoretical ratios like those illustrated
in Figure~\ref{f:linerats}.   In these calculations, specific account of the 
possible contribution of Fe~XIX blends at 14.423 and 14.462~\AA\ 
to the Ne~XI flux was taken.  The largest contribution found amounted 
to 15\%,  but was less than 10\%\ for the great majority of the lines, 
confirming the assessment of \citet{McKenzie.Feldman:92} that
Fe~XIX blends are not significant for these observations.
The abundance ratios as a function of temperature for both isothermal and DEM cases are shown in
Figure~\ref{f:abuns}, together with their error-weighted linear
best fits.  Best-fits accounted for errors both in the Ne/O flux ratios,
and in the temperatures derived from the Fe~XVII and Fe~XVIII lines.

The temperature dependence in the derived Ne/O
abundance ratios is again obvious, with both DEM and isothermal cases
showing a rise to higher Ne/O with rising temperature.  The mean
Ne/O ratio is slightly lower than values proposed by
\citet[Ne/O=0.15 by number, or $-0.82$ in logarithm]{Asplund.etal:05},  \citet[Ne/O=0.17 by number, or $-0.76$ in log]{Asplund.etal:09} 
and \citet[Ne/O=0.18 by number or $-0.75$ in log]{Grevesse.Sauval:98}.  
The data of \citet{Schmelz.etal:05b} did not include the Fe fluxes we require here to derive DEM peak temperatures and are therefore not shown.   However, it is clear from the overlap of \citet{Schmelz.etal:05b} and \citet{McKenzie.Feldman:92} line intensity ratios shown in Figure~\ref{f:linerats}, and comparison of Figures~\ref{f:linerats} and \ref{f:abuns}, that those data
also correspond to a mean slightly lower than these values.  


\section{Discussion}
\label{s:discuss}

\subsection{Atomic Data}
\label{s:atomic}

Can errors in the underlying atomic data be responsible for the
apparent trend of Ne/O with temperature?   The He-like and H-like ions
present the simplest cases for computing both ion balances and
collisional excitation rates, and atomic data for these species should
in general be more accurate than for more complex ions.  The shape of
the theoretical Ne~IX/O~VIII emissivity ratio in the temperature range
of interest here is determined primarily by the increasing excitation
rate with temperature for both species and the ramp-down of the Ne~IX
and O~VIII ion populations toward higher temperatures.  It is
difficult to imagine incurring an error of a factor of $\sim 2$ in
the ratio of the former.   Revised assessments of the emissivity 
ingredients for the Ne~IX and 
O~VIII resonance lines were included in version 6 of the CHIANTI database 
 \citep{Dere.etal:09} used here.   \citet{Dere.etal:09} note that O~VIII line intensities are 
in close agreement with the previous ones.   \citet{Chen.etal:06} also reported  $R$-Matrix 
collision strengths for Ne~IX within a few percent of early data.  We have nevertheless  repeated the calculations reported here for CHIANTI versions 4 and 5, and find only very small differences of less than 10\%\ in derived abundances.

The rising Ne abundance with increasing
temperature observed in active regions could be mimiced by an ion
population error in Ne~IX, but the required errors are again of order
a factor of 2 at temperatures of $\log T\sim 6.6$, or a shift in the Ne~IX ion population curve toward higher
temperatures by 25\%\ or so.  Such a shift would also require
commensurate changes in the Ne~X population, and it seems difficult to
introduce such a change in Ne without invoking similar changes in the
O ion balance.  The calculations presented here employ the most 
recent ion balance assessment currently available \citep{Bryans.etal:09},
and it seems highly unlikely that residual errors of such magnitude 
remain.  We have also repeated the abundance calculations for the 
ionization equilibria of \cite{Mazzotta.etal:98} and \citet{Arnaud.Rothenflug:85}.
The latter result in slightly lower temperatures from Fe~XVII and Fe~XVIII lines
by about 0.1~dex, and systematically higher Ne/O abundance ratios by
a similar amount, but the abundance trends are unaffected (see also \S\ref{s:saturated} below).

\citet{Desai.etal:05} found the Fe~XVIII~$\lambda 14.21$ resonance line strength predicted by CHIANTI v4.2 and other databases appeared lower by 25\%\ relative to the 
$\lambda 93.92$ resonance $2p-2s$ line compared with the ratio observed in
Capella.   More recent Fe~XVIII electron 
collisional excitation calculations have been published by \citet{Witthoeft.etal:07}.  Using these data, 
\citet{Del_Zanna:06} finds good agreement between observed and predicted Fe~XVIII line strengths.  It is these data that are included in CHIANTI v6.0.1 used here.  

We conclude that atomic data errors are a very unlikely
explanation for the observed temperature trend in Ne/O, and that it is
the underlying abundance ratio itself which is responsible.

\subsection{A ``Neon Saturation" interpretation of fractionation in the 
coronae of the Sun and stars}
\label{s:saturated}

We noted in \S\ref{s:abunrats} that the mean Ne/O abundance ratio
indicated by the data in Figure~\ref{f:abuns} is somewhat lower than
the assessments of \citet{Asplund.etal:05,Asplund.etal:09} and
\citet{Grevesse.Sauval:98}.  The mean ratio would tend to
reconciliation with these values were the estimated active region
temperatures to be systematically too high by $\sim 0.15$~dex or so.
In this regard, the ionization balance of \citet{Arnaud.Rothenflug:85} provides 
better agreement with the solar assessments than that of \citet{Bryans.etal:09}.
However, the trend with temperature cannot be erased by plausible adjustments
of the estimated temperatures: the scatter in the observed Ne/O ratios
is simply much larger than variations in the theoretical line ratio.  The
main result we emphasise here is this temperature dependence in the
abundance ratio, rather than its absolute value.  

By comparison of Fe/O and Ne/O ratios, \citet{McKenzie.Feldman:92}
showed that it is the abundance of Ne that varies rather than that of
O.  This is also expected on other grounds: the almost identical
ionization potentials of neutral O and H, and the consequently large
charge-exchange cross-section between their neutral and ionized
species, should couple these elements quite efficiently such
that the coronal O abundance is expected to follow that of H.  Echoing
\citet{McKenzie.Feldman:92}, we conclude that the coronal Ne abundance
varies by more than a factor of 2.  We also conclude here that the
mechanism that fractionates Ne operates so as to either enhance hotter
plasma with Ne, or to deplete Ne in cooler plasma (or perhaps both).
Until this fractionation mechanism can be identified and understood,
it is not obvious which case applies.  This has profound consequences
for the solar Ne abundance since it is not yet possible to ascertain
whether the cooler or hotter regions represent the true Ne abundance,
or {\em whether any region of the solar outer atmosphere has a Ne
content representative of that of the bulk of the Sun}.  The
well-known chemical fractionation based on first ionization potential (FIP)
observed in the solar corona is thought to originate in the
chromosphere \citep[e.g.\ ][]{Meyer:85,Feldman:92}.  If the Ne
fractionation operates in a similar region, it is then quite possible
that the Ne abundance is modified in {\em all} regions of the corona,
and in the solar wind and energetic particles.   Assessment of the true solar Ne content must then be undertaken deeper into the atmosphere \citep[see][for a method based on photospheric X-ray fluorescence]{Drake.Ercolano:07}.


Based on the constancy of the Ne/O abundance ratio derived from {\it
Chandra} X-ray spectra of a sample of mostly quite magnetically active
stars, \citet{Drake.Testa:05} concluded that the data likely
represented the true underlying stellar Ne/O abundance ratios.  Their
value is $\sim 2.7$ times higher than that of \citet{Asplund.etal:05}, albeit with
some uncertainty, and 2.4 times higher than the \citet{Asplund.etal:09} ratio.  An 
assessment of Ne/O ratios for a larger stellar sample culled from the literature by 
\citet{Garcia-Alvarez.etal:09} finds a slightly lower factors of 2.1 and 1.9, respectively, or Ne/0$=0.32\pm^{0.14}_{0.10}$.
In this regard, the increasing Ne/O ratio we see toward hotter coronal
temperatures is quite conspicuous.  Using plasma temperature as an 
activity proxy, the Ne/O ratio increases as a function of solar region ``activity". 
This activity-dependent Ne content of the solar corona fits in well with a recent study of Ne/O ratios in low-activity stars by \citet{Robrade.etal:08}, who find evidence for a 
trend of higher Ne/O ratios with increasing stellar activity level (but assumed the lower stellar activities represented photospheric abundances).  A similar trend is also possibly present in the compilation of \citet{Gudel:04}.   


Based on existing results, we echo the earlier conclusions of  \citet{Drake.Testa:05} and suggest that active stars represent coronal 
``neon saturation", with Ne content reaching photospheric levels.   It seems otherwise unlikely that the Ne/O ratio can be so well controlled by a chemical fractionation mechanism when in the same stars fractionation varies the Fe/O ratio by an order of 
magnitude \citep[e.g.][]{Gudel:04,Garcia-Alvarez.etal:08,Garcia-Alvarez.etal:09}.  
In this scenario, the solar corona, and the coronae of similar ``low activity'' stars, is 
{\em depleted} in Ne by factors $\sim 1.5-4$; the currently
recommended solar Ne abundance underestimates the true abundance by
a factor of about 2, or possibly more.  A ratio Ne/O$=0.32\pm^{0.14}_{0.10}$ is also formally consistent with the valued $0.19\pm_{0.06}^{0.10}$ preferred by \citet{Delahaye.etal:10} based on helioseismology constraints.

This picture of depleted Ne in the solar corona is not unprecedented: the element with 
a FIP closest to that of Ne is He, which is depleted in the solar corona and wind by a factor of 2 or so \citep[e.g.][and references therein]{Laming:09}.  \citet{Laming:09} presents a fractionation model based on the ponderomotive force resulting from the oscillating electric field of Alfv\'en waves propagating through the chromosphere and corona.  This model predicts coronal depletion of He to the different degrees required by observations of slow and fast solar winds, and also of Ne by a factor of $\sim 2$.

Neon to oxygen ``overabundance" ranging up to factors of 2 or more 
relative to the old canonical Ne/O=0.15 mixture 
is also the prevalent situation in many representative objects of the nearby Galaxy, including B stars and BA-type supergiants  \citep{Kilian:94,Kilian-Montenbruck.etal:94,Sigut:99,
Przybilla.etal:06,Cunha.etal:06,Morel.Butler:08b,Lanz.etal:08}, and planetary nebulae and H~II
regions \cite[e.g.][]{Tsamis.etal:03,Liu.etal:04,Perinotto.etal:04, Pottasch.Bernard-Salas:06, Stanghellini.etal:06,Wang.Liu:08,Magrini.etal:09,Rubin.etal:11}.   Upward revision of the solar Ne/O ratio by a factor of 2 would not be out of line with some of these findings.   


\section{Conclusions}
\label{s:con}

An analysis using modern atomic data of solar active region X-ray spectra obtained in the 1970s and 1980s confirms that 
the coronal Ne content varies by a factor of 2 or more, and reveals 
a trend of increasing Ne abundance with increasing plasma
temperature.  This trend seems to reflect the emerging picture of 
Ne/O abundances in late-type stellar coronae, with Ne/O appearing to increase 
with stellar activity until quickly reaching a fairly constant ``neon saturation" in only moderately active stars.   We argue
this latter value represents photospheric ratios.  Instead, the Sun is more capricious and withholds neon (as well as helium), from its outer atmosphere: the true solar Ne abundance cannot yet be inferred with
any degree of certainty from any existing observations of the outer
solar atmosphere or wind.  Under the ``neon saturation" hypothesis it is 
more abundant than currently assessed, perhaps by as much as a factor of two.


\acknowledgments

The author is gratefu to the anonymous referee for helping clarify and improve the manuscript.  JJD thanks the NASA AISRP for providing financial assistance for the
development of the PINTofALE package.  JJD was funded by NASA contract
NAS8-39073 to the {\em Chandra X-ray Center} during the course of this
research and thanks the Director, H.~Tananbaum, for continuing support
and encouragement.



\newpage

\end{document}